\documentclass[10pt,twoside]{article}
\usepackage{fancyhdr}
\usepackage{titlesec}
\usepackage{cite}
\usepackage{ifthen}
\usepackage{graphicx}
\usepackage{float}

\titleformat{\section}{\centering\large\bfseries}{\S\arabic{section}}{1em}{}
\newboolean{first}
\setboolean{first}{true}

\begin{document}

\setlength\abovedisplayskip{2pt}
\setlength\abovedisplayshortskip{0pt}
\setlength\belowdisplayskip{2pt}
\setlength\belowdisplayshortskip{0pt}

\title{\bf \Large  A secure and effective anonymous authentication scheme for roaming service in global mobility networks
\author{Dawei Zhao$^{a,b}$  \  \ Haipeng Peng$^{a,b}$   \  \ Lixiang Li$^{a,b}$ \ \ Yixian Yang$^{a,b}$ \\ \small \it $^{a}$Information Security Center, Beijing University of Posts and
Telecommunications, \\
\small \it Beijing 100876, China. \\ \small \it $^{b}$National
Engineering Laboratory for Disaster Backup and
Recovery, \\
\small \it Beijing University of Posts and Telecommunications,
Beijing 100876, China. }\date{}} \maketitle

\footnote{E-mail address: dwzhao@ymail.com (Dawei Zhao);
penghaipeng@bupt.edu.cn (Haipeng Peng).}

\begin{center}
\begin{minipage}{135mm}
{\bf \small Abstract}.\hskip 2mm {\small Recently, Mun et al.
analyzed Wu et al.'s authentication scheme and proposed a enhanced
anonymous authentication scheme for roaming service in global
mobility networks. However, through careful analysis, we find that
Mun et al.'s scheme is vulnerable to impersonation attacks, off-line
password guessing attacks and insider attacks, and cannot provide
user friendliness, user's anonymity, proper mutual authentication
and local verification. To remedy these weaknesses, in this paper we
propose a novel anonymous authentication scheme for roaming service
in global mobility networks. Security and performance analyses show
the proposed scheme is more suitable for the low-power and
resource-limited mobile devices, and is secure against various
attacks and has many excellent features.}
\end{minipage}\end{center}
\begin{center}
\begin{minipage}{135mm}
\end{minipage}
\end{center}

\section{Introduction}
\label{}

GLOBAL mobility network (GLOMONET) [1] provides global roaming
service that permits mobile user to use the services provided by
his/her home agent ($HA$) in a foreign agent ($FA$). When a mobile
user roams into a foreign network, mutual authentication must first
be solved to prevent illegal use from accessing services and to
ensure that mobile users are connected to a trusted networks. A
strong user authentication scheme in GLOMONET should satisfy the
following requirements: (1) user anonymity; (2) low communication
cost and computation complexity; (3) single registration; (4) update
session key periodically; (5) user friendly; (6) no
password/verifier table; (7) update password securely and freely;
(8) prevention of fraud; (9) prevention of replay attack; (10)
security; and (11) providing the authentication scheme when a user
is located in the home network. More details about these
requirements can be found in [2].

In order to achieve secure and effective mutual authentication and
privacy protection in GLOM-ONET, many authentication protocols have
been proposed [2-17]. In 2004, Zhu and Ma [3] proposed an
authentication scheme with anonymity for wireless environments.
However, Lee et al. [4] pointed out that Zhu et al.'s scheme [3]
cannot achieve mutual authentication and perfect backward secrecy,
and is vulnerable to the forgery attack. At the same time, Lee et
al. proposed an enhanced anonymous authentication scheme, but Chang
et al. [5] and Wu et al. [6] found that Lee et al.'s scheme also
cannot achieve user'anonymity, and an attacker who has registered as
a user of an $HA$ can obtain the identity of other users as long as
they registered at the same $HA$. After that, in 2011, Li et al. [2]
found Wu et al. [6] is unlikely to provide user's anonymity due to
an inherent design weakness and also vulnerable to replay and
impersonation attacks. Then they constructed a strong user
authentication scheme with smart cards for wireless communications.
However, Li and Lee [7] showed that Li et al.'s scheme [2] lacks of
user friendliness, and cannot provide user's anonymity and
unfairness in key agreement.

Recently, Mun et al. [8] reanalyzed Wu et al.' authentication scheme
[6], they point out that Wu et al.'s scheme also fails to achieve
user's anonymity and perfect forward secrecy, and discloses of
legitimate user's password. Then they proposed an enhanced anonymous
authentication scheme for roaming service in global mobility
networks. However, through careful analysis, we find that Mun et
al.'s scheme is vulnerable to impersonation attacks, off-line
password guessing attacks and insider attacks, and cannot provide
user friendliness, user's anonymity, proper mutual authentication
and local verification. To remedy these weaknesses, in this paper we
propose a novel anonymous authentication scheme for roaming service
in global mobility networks. Security and performance analyses show
the proposed scheme is more suitable for the low-power and
resource-limited mobile devices, and is secure against various
attacks and has many excellent features.

The remainder of this paper is organized as follows. Section 2
provides some basic knowledge. In Section 3, we review Mun et al.'s
scheme and Section 4 shows the security weaknesses of Mun et al.'s
scheme. A novel user authentication scheme is proposed in Section 5.
In Section 6, we analyze the security of our proposed scheme. Next,
we compare the functionality and performance of our proposed scheme
and make comparisons with other related schemes in Section 7.
Finally, in Section 8 we make some conclusions.

\section{Preliminaries}

In this section, we briefly introduce the elliptic curve
cryptosystem and some related mathematical assumptions. Compared
with other public key cryptography, elliptic curve cryptosystem
(ECC) has significant advantages like smaller key sizes, faster
computation. It has been widely used in several cryptographic
schemes of wireless network environment to provide desired level of
security and computational efficiency.

\subsection{Elliptic curve cryptosystem}

Let $E_p(a,b)$ be a set of elliptic curve points over the prime
field $E_p$, defined by the non-singular elliptic curve equation:
$y^2$mod $p$ $=(x^3 + ax + b)$mod$p$ with $a, b\in F_p$ and $(4a^3 +
27b^2)$mod$p\neq 0$. The additive elliptic curve group defined as
$G_p = \{(x, y): x, y\in F_p$ and $(x, y)\in E_p(a, b)\}\cup {O}$,
where the point $O$ is known as ``point at infinity". The scalar
multiplication on the cyclic group $G_p$ defined as $k\cdot P =
\underbrace{P + P + \cdot\cdot\cdot + P}\limits_{k\
\emph{\emph{times}}}$. A point $P$ has order $n$ if $n\cdot P = O$
for smallest integer $n> 0$. More details about elliptic curve group
properties can be found in [18-20].

\subsection{Related mathematical assumptions} \label{}

To prove the security of our proposed protocol, we present some
important computational problems over the elliptic curve group which
are frequently used to design secure cryptographic schemes.

(1) Computational discrete logarithm (CDL) problem: Given $R =
x\cdot P$, where $P, R \in  G_p$. It is easy to calculate $R$ given
$x$ and $P$, but it is hard to determine $x$ given $P$ and $R$.

(2) Computational Diffie-Hellman (CDH) problem: Given $P, xP, yP \in
G_p$, it is hard to compute $xyP\in  G_p$.

(3) Elliptic curve factorization (ECF) problem: Given two points $P$
and $R= x\cdot P +y\cdot P$ for $x, y\in Z^*_q$ , it is hard to find
$x\cdot P$ and $y\cdot P$.

\section{Review of Mun et al.'s scheme} \label{}

In this section, we briefly review the Mun et al.'s scheme [8].
There are three phases in their scheme: registration phase,
authentication and establishment of session key phase, and update
session key phase. Three entities are involved: $MU$ is a mobile
user, $FA$ is the agent of the foreign network, and $HA$ is the home
agent of the mobile user $MU$. Table 1 lists some notations used in
Mun et al.¡¯s scheme.

\begin{table}\centering{
\caption{Notations used in Mun et al.'s scheme.}
\label{tab:1}       % Give a unique label
% For LaTeX tables use
\begin{tabular}{lllll}
\hline\noalign{\smallskip} Notation & Description\\
\hline\noalign{\smallskip} $MU$, $FA$, $HA$& Mobile User, Foreign
Agent, Home Agent\\
$PW_X$& Password of an entity $X$\\ $ID_X$& Identity of an entity
$X$\\ $h(\cdot)$& A one-way hash function\\ $N_X$& Number used only
once (Random number) generated by an entity $X$\\ $\|$& Concatenation operation\\
$\oplus$& XOR operation\\ $f_K$& MAC generation function by using
the key $K$\\ $K_{XY}$& Session key between entity $X$ and $Y$\\
\noalign{\smallskip}\hline
\end{tabular}}
% Or use
%\vspace*{5cm}  % with the correct table height
\end{table}

\subsection{Registration phase} \label{}
When a mobile user $MU$ wants to become a legal client to access the
services, $MU$ needs to register himself/herself to his/her home
agent $HA$. The handshake between $MU$ and $HA$ is depicted in Fig.
1.

\begin{figure}[h]
\centering{\includegraphics[scale=0.7,trim=0 0 0 0]{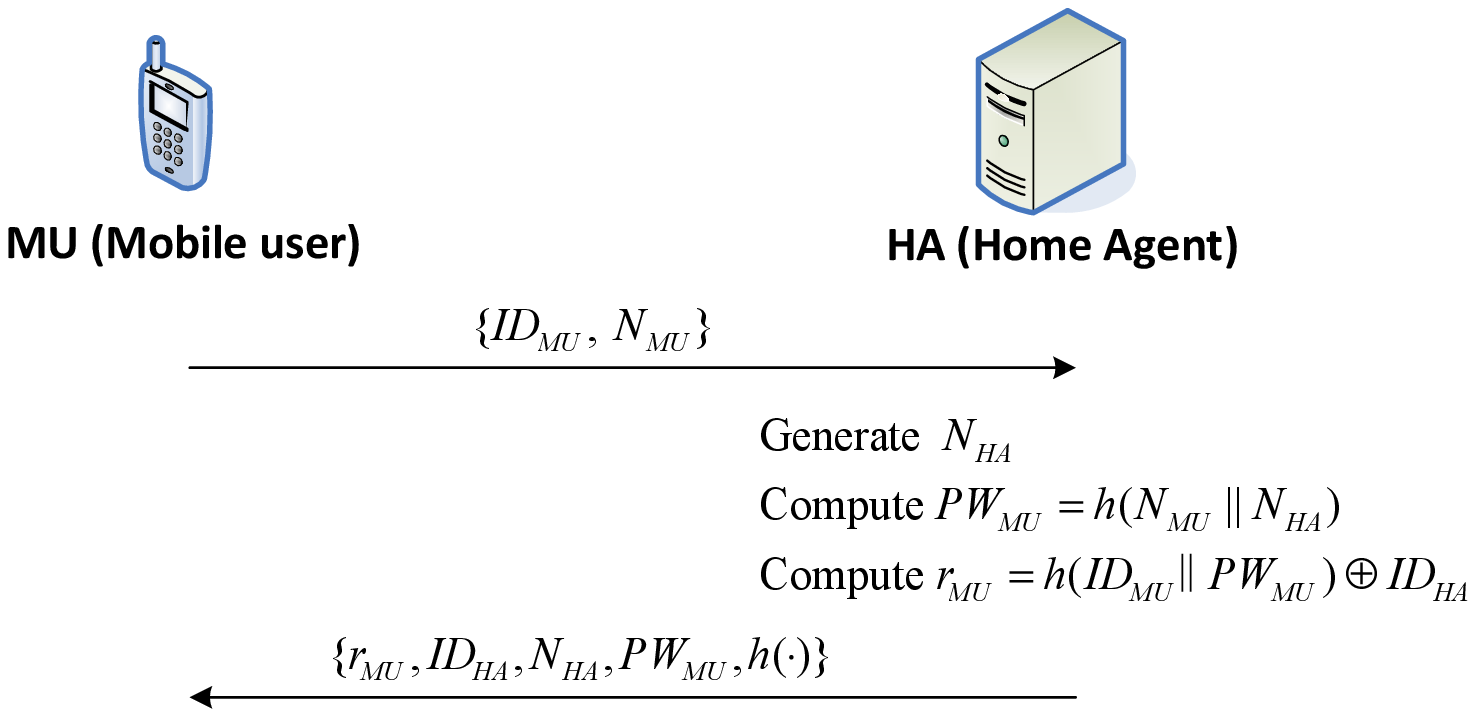}}
\caption{Registration phase of Mun et al.'s scheme.}
\end{figure}

\textbf{Step R1}: $MU$ sends his/her identity $ID_{MU}$ and a random
number $N_{MU}$ to $HA$.

\textbf{Step R2}: $HA$ generates a random number $N_{HA}$ and
computes $PW_{MU} = h(N_{MU}\| N_{HA})$ and $r_{MU} = h(ID_{MU}\|
PW_{MU})\oplus ID_{HA}$.

\textbf{Step R3}: $HA$ sends $r_{MU}$, $PW_{MU}, N_{HA}, ID_{HA},$
and $h(\cdot)$ to $MU$ through a secure channel.

\subsection{Authentication and establishment of session key phase} \label{}
When a mobile user $MU$ roams into a foreign network $FA$ and wants
to access services provided by $FA$. The $FA$ needs to verify the
validity of $MU$ with the assistance of $HA$, and proves to $MU$
that he is a legitimate service provider. The authentication and
establishment of session key phase of Mun et al.'s scheme is shown
in Fig.2.

\begin{figure}[h]
\centering{\includegraphics[scale=0.7,trim=0 0 0 0]{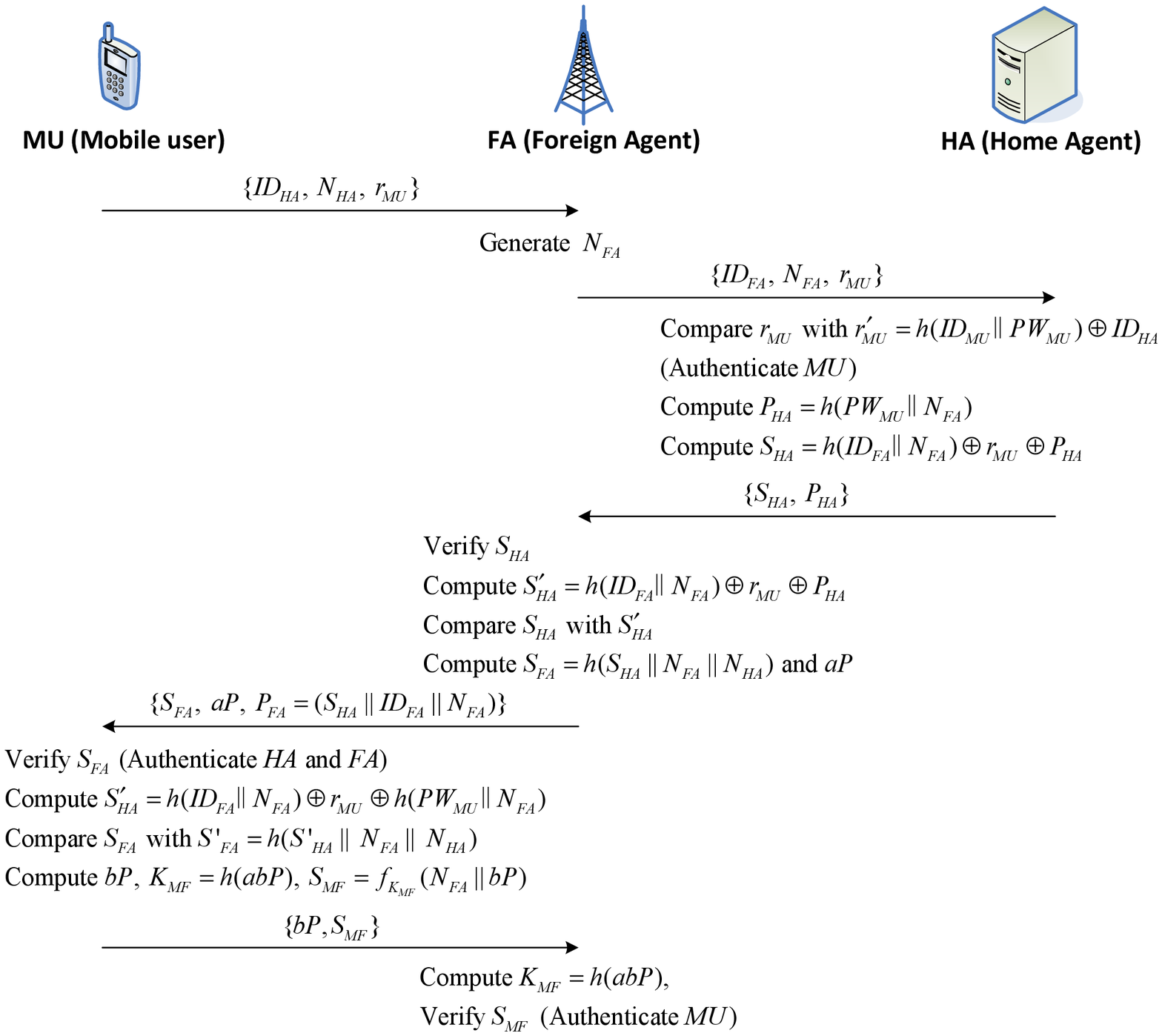}}
\caption{Authentication and establishment of session key phase of
Mun et al.'s scheme.}
\end{figure}

\textbf{Step A1}: $MU$ submits $ID_{HA}$, $N_{HA}$ and $r_{MU}$ to
$FA$.

\textbf{Step A2}: $FA$ stores the received message from $MU$ for
further communications and generates a random number $N_{FA}$. Then,
$FA$ sends $ID_{FA}$, $N_{FA}$ and $r_{MU}$ to $HA$.

\textbf{Step A3}: After receiving the message sent from $FA$, $HA$
computes $r'_{MU} = h(ID_{MU}\| PW_{MU})$ $\oplus ID_{HA}$ and
compares it with the received $r_{MU}$. If they are not equal, $HA$
considers $MU$ as illegal user and terminates this procedure.
Otherwise, $HA$ can authenticate $MU$. Next, $HA$ computes $P_{HA} =
h(PW_{MU}\| N_{FA})$ and $S_{HA} = h(ID_{FA}\| N_{FA})\oplus
r_{MU}\oplus P_{HA}$. Then, $HA$ sends the computed $S_{HA}$ and
$P_{HA}$ to $FA$.

\textbf{Step A4}: When receiving $S_{HA}$ and $P_{HA}$ sent from
$HA$, $FA$ computes $S'_{HA} = h(ID_{FA}\| N_{FA})$ $\oplus
r_{MU}\oplus P_{HA}$ and . Then, $FA$ verifies whether $S'_{HA}$
equals the received $S_{HA}$. If the result is not correct, the
procedure is terminated. Next, $FA$ computes $S_{FA} = h(S_{HA}\|
N_{FA}\| N_{HA})$, selects random number $a$, and computes $aP$ on
$E$ by using Elliptic Curve Diffie-Hellman (ECDH) []. After that,
$FA$ sends $S_{FA}$, $aP$ and $P_{FA} = (S_{HA}\| ID_{FA}\| N_{FA})$
to $MU$.

\textbf{Step A5}: First, $MU$ computes $S'_{HA} = h(ID_{FA}\|
N_{FA})\oplus r_{MU}\oplus h(PW_{MU}\| N_{FA})$ and $S'_{FA} =
h(S'_{HA}\| N_{FA}\| N_{HA})$. Then, $MU$ checks whether
$S'_{FA}=S_{FA}$. If they are not equal, the procedure is
terminated. Otherwise, $MU$ can authenticate $FA$ and $HA$.
Afterwards, $MU$ selects a random number $b$, and computes $bP$ and
a session key $K_{MF} = h(abP)$. Moreover, $MU$ computes $S_{MF} =
f_{K_{MF}} (N_{FA}\| bP)$, and sends $bP$ and $S_{MF}$ to $FA$.

\textbf{Step A6}: After receiving the message sent from $MU$, $FA$
computes $K_{MF} = h(abP)$ and $S'_{MF} = f_{K_{MF}} (N_{FA}\| bP)$.
FA verifies whether $S'_{MF}$ equals the received $S_{MF}$. If the
result is not correct, session key $K_{MF} = h(abP)$ between $MU$
and $FA$ is not valid and $FA$ terminates the procedure. Otherwise,
$FA$ can authenticate $MU$.

\subsection{Update session key phase} \label{}

$MU$ and $FA$ need to renew session key for security reasons if user
is always within a same $FA$. When $MU$ visits $FA$ at the $i$th
session, the following process is conducted to authenticate $FA$:

\textbf{Step U1}: $MU$ selects a new random number $b_i$, computes
$b_iP$ $(i = 1, 2, . . . , n)$, and sends $b_iP$ to FA.

\textbf{Step U2}: $FA$ selects a new random number $a_i$ and
computes $a_iP$ $(i = 1, 2, . . . , n)$. Then $FA$ generates a new
session key $K_{MF_i} = h(a_ib_iP)$, and then computes $S_{MF_i} =
f_{K_{MF_i}} (a_ib_iP\| a_{i-1}b_{i-1}P)$. After that, $FA$ sends
$a_iP$ and $S_{MF_i}$ to $MU$.

\textbf{Step U3}: $MU$ computes session key $K_{MF_i} = h(a_ib_iP)$
by using the received $a_iP$. $MU$ computes $S'_{MF_i} =
f_{K_{MF_i}} (a_ib_iP\| a_{i-1}b_{i-1}P)$. Then, $MU$ checks whether
$S'_{MF_i}=S_{MF_i}$. If they are equal, the new session key
$K_{MF_i} = h(a_ib_iP)$ is established between $MU$ and $FA$.

Procedure of update session key phase is depicted in Fig.3.

\begin{figure}[h]
\centering{\includegraphics[scale=0.7,trim=0 0 0 0]{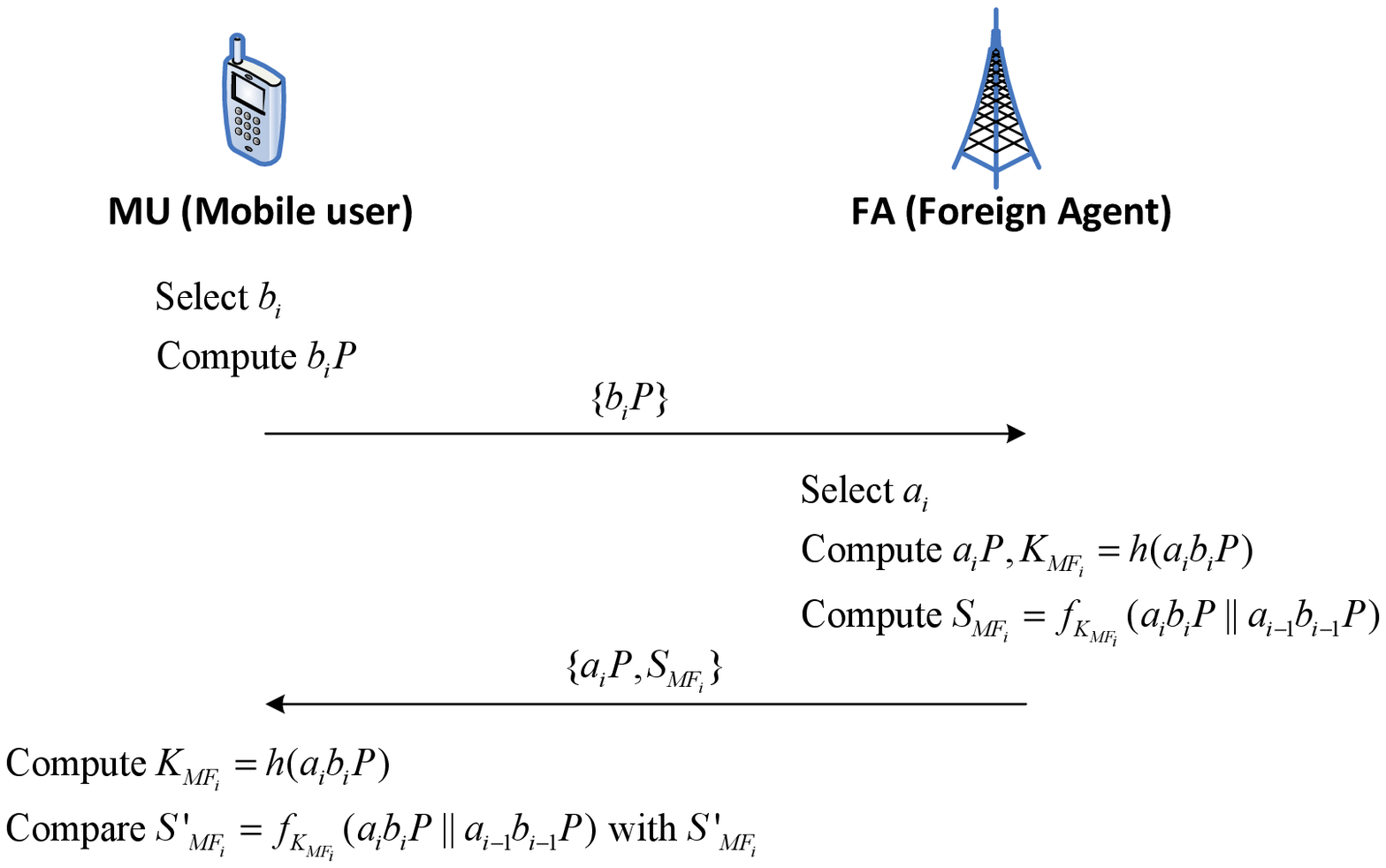}}
\caption{Update session key phase of Mun et al.'s scheme.}
\end{figure}

\section{Weaknesses of Mun et al.'s scheme} \label{}

Recently, Kim and Kwak [21] pointed out that Mun et al.'s scheme [8]
cannot withstand replay attacks and man-in-the-middle attacks.
Through careful analysis, in this section we show that Mun et al.'s
scheme is also vulnerable to impersonation attacks, off-line
password guessing attacks and insider attacks, and cannot provide
user friendliness, user's anonymity, proper mutual authentication
and local verification.

\subsection{Impersonation attacks}

\subsubsection{$MU$ impersonation attacks}

In Mun et al.'s scheme, an attacker can masquerade as a user $MU$ to
cheating any foreign agent $FA'$ and $MU$'s $HA$ if he/she has
intercepted a valid login request message $\{ID_{HA}, N_{HA},
r_{MU}\}$ of $MU$. First, the attacker generates a random number
$N'_{HA}$ and sends $\{ID_{HA}, N'_{HA}, r_{MU}\}$ to $FA'$. Since
$ID_{HA}$ and $r_{MU}$ are the real home agent and correct personal
information of $MU$ respectively, the login request message can pass
the validation of $HA$. Furthermore, $HA$ will notify the $FA'$ that
the attacker who is masquerading as the user $MU$ is a legitimate
user. Therefore, the attacker can further establish a session key
with $FA'$ and access the services provided by $FA'$.

\subsubsection{$FA$ impersonation attacks}

In the authentication and establishment of session key phase of Mun
et al.'s scheme, it can be found that the $HA$ only authenticates
the $MU$ by verifying the received $r_{MU}$ but do not make any
authentication to the $FA$. At the same time, there is no secret
information of $FA$ in the message $\{ID_{FA}, N_{FA}, r_{MU}\}$
sent from $FA$ to $MU$'s $HA$. Thus an attacker can masquerade as a
foreign agent $FA$ to cheating any user $MU'$ and $MU'$'s $HA$. For
example, if the attacker intercepts a login request message
$\{ID_{HA}, N_{HA}, r_{MU'}\}$ sent from $MU'$ to $FA$, the attacker
can generate a random number $N_{FA}$ and send $\{ID_{FA}, N_{FA},
r_{MU'}\}$ to $HA$ by masquerading as $FA$. Since $r_{MU'}$ is the
correct personal information of $MU'$ and there is no identity
authentication process of $HA$ to $FA$. Therefore, the message
$\{ID_{HA}, N_{HA}, r_{MU'}\}$ can pass the authentication of $HA$.
At the same time, since the authentication of $MU'$ to $FA$ is
completely dependent on $HA$ and $FA$ has been authenticated by
$HA$, the $FA$ will pass the authentication of $MU'$. Therefore, the
attacker who is masquerading as the $FA$ can establish a session key
with $MU'$ and tricks $MU'$ successfully.

\subsubsection{$HA$ impersonation attacks}

In the authentication and establishment of session key phase of Mun
et al.'s scheme, the $FA$ authenticates $MU$ and $HA$ by verifying
whether $S'_{HA}=S_{HA}$. However, there is a security vulnerability
in this step such that an attacker can masquerade as a home agent to
help any agent pass the authentication of a $FA$ and access the
services provided by $FA$. It is assumed that $B$ is a agent who
wants to access the services provided by $FA$ and $A$ is an attacker
who masquerades as $B$'s home agent $HA$ to help $B$ pass the
authentication of $FA$.

First, $B$ freely chooses two numbers $N'$ and $r'$, and submits
$\{ID_{HA}, N', r'\}$ to $FA$. Then $FA$ generates a random number
$N_{FA}$ and sends the message $\{ID_{FA}, N_{FA}, r'\}$ to $HA$.
Right now, $A$ intercepts this message, freely chooses a number
$P'$, and computes $S_{HA} = h(ID_{FA}\| N_{FA})\oplus r' \oplus
P'$. Then, $A$ sends the computed $S_{HA}$ and $P'$ to $FA$. When
receiving $S_{HA}$ and $P'$ sent from $A$ who is masquerading as the
$HA$, $FA$ computes $S'_{HA} = h(ID_{FA}\| N_{FA})\oplus r'\oplus
P'$. Obviously, the $S'_{HA}$ equals the received $S_{HA}$. Next,
$FA$ computes $S_{FA} = h(S_{HA}\| N_{FA}\| N')$, selects random
number $a$, and computes $aP$. After that, $FA$ sends $\{S_{FA}, aP,
P_{FA} = (S_{HA}\| ID_{FA}\| N_{FA})\}$ to $B$. At this point, $B$
does not need to verify the $S_{FA}$, but directly chooses a random
number $b$ and computes $K_{MF} = h(abP)$ and $S_{MF} = f_{K_{MF}}
(N_{FA}\| bP)$. Then $B$ sends $bP$ and $S_{MF}$ to $FA$. After
receiving $\{bP, S_{MF}\}$ sent from $B$, $FA$ computes $K_{MF} =
h(abP)$ and $S'_{MF} = f_{K_{MF}} (N_{FA}\| bP)$. Obviously, this is
$S'_{MF}=S_{MF}$. $FA$ thus authenticates $B$. By the above method,
with the assistance of $A$, $B$ establishes the session key $K_{MF}
= h(abP)$ with $FA$ and can access the services provided by $FA$.

\subsection{Off-line password guessing attacks}

Most passwords have such low entropy that it is vulnerable to
password guessing attacks, where an attacker intercepts useful
information from the open channel or the lost smart card. In Mun et
al.'s scheme, an attacker is assumed to have intercepted a previous
full transmitted messages $\{ID_{HA}$, $N_{HA}$, $r_{MU}$,
$ID_{FA}$, $N_{FA}$,$r_{MU}$, $S_{HA}$, $P_{HA}$, $S_{FA}$, $aP$,
$P_{FA} = (S_{HA}\| ID_{FA}\| N_{FA})$, $bP$, $S_{MF}\}$. The
attacker can submit the guessing password $PW'_{MU}$ and compute
$S'_{HA} = h(ID_{FA}\| N_{FA})\oplus r_{MU}\oplus h(PW'_{MU}
\parallel N_{FA})$. If the computed $S'_{HA}$ is equal to $S_{HA}$,
the attacker can regard the guessing password $PW'_{MU}$ as the
original password $PW_{MU}$. Therefore, Mun et al.'s scheme cannot
withstand the off-line password guessing attacks.

\subsection{Insider attacks}

In the registration phase, $MU$ sends $ID_{MU}$ and a random number
$N_{MU}$ to $HA$. Then $HA$ generates a random number $N_{HA}$,
computes $PW_{MU} = h(N_{MU}\| N_{HA})$ and $r_{MU} = h(ID_{MU}\|
PW_{MU}) \oplus ID_{HA}$, and sends $\{r_{MU}, PW_{MU}, N_{HA},
ID_{HA}, h(\cdot)\}$ to $MU$ through a secure channel. It is obvious
that the $HA$ knows all the secret information of $MU$ so that $HA$
can impersonate $MU$ to do anything. Therefore, Mun et al.'s scheme
is vulnerable to the insider attack.

\subsection{Lack of user friendliness}

User friendliness means that the proposed authentication scheme
should be easily used by users. However, in the registration phase
of Mun et al.'s scheme, the home agent $HA$ sends the information
$\{r_{MU}$, $PW_{MU}, N_{HA}, ID_{HA}, h(\cdot)\}$ to the user $MU$
without using smart card. So that $MU$ needs to remember and enter
so much information in the authentication and establishment of
session key phase. Therefore, Mun et al.'s scheme is actually
infeasible and unrealistic.

\subsection{Lack of user's anonymity}

In the second phase of Mun et al.'s scheme, $MU$ sends $r_{MU}$ to
$FA$ instead of his/her real identity $ID_{MU}$. Thus the authors
claimed that their scheme achieves the user's anonymity. However, in
each login message $\{ID_{HA}, N_{HA}, r_{MU}\}$ of $MU$, the
contents of $N_{HA}$ and $r_{MU}$ are always unchanged. Any attacker
could easily trace $MU$ according to $N_{HA}$ and $r_{MU}$ and thus
the user's anonymity cannot achieved.

\subsection{Lack of proper mutual authentication}

In Mun et al.'s scheme, the $HA$ does not maintain any verification
table. Thus after receiving the message $\{ID_{FA}, N_{FA},
r_{MU}\}$ sent from $FA$, $HA$ cannot recognize which user launched
the authentication request to $FA$. So $HA$ cannot computes
$r'_{MU}$ and checks it with the received $r_{MU}$. On the other
hand, even if $HA$ can compute $r'_{MU}$ and check whether
$r'_{MU}=r_{MU}$, it only means $HA$ authenticates the legality of
$MU$. However, it is found that $HA$ do not make any authentication
to the $FA$. Therefore, Mun et al.'s scheme cannot provide proper
mutual authentication.

\subsection{Lack of local verification}

In the authentication and establishment of session key phase of Mun
et al.'s scheme, the $MU$ directly enters and sends the login
message to $FA$. Note that the smart terminal of $MU$ does not
verify the entered information correctly or not. Therefore, even if
the $MU$ enters the login message incorrectly by mistake or an
attacker sends an forged message, the authentication phase still
continue in their scheme. This obviously results to cause
unnecessarily having extra communication and computational costs.

\section{The proposed scheme}
\label{}

In this section, we propose a novel anonymous authentication scheme
for roaming service in global mobility networks using elliptic curve
cryptosystem to not only protect the scheme from security breaches,
but also emphasize the efficient features. In addition to including
the general registration phase, authentication and establishment of
session key phase and update session key phase, our scheme also
contains the update password phase and authentication and
establishment of session key scheme when a mobile user is located in
his/her home network. Table 2 lists some notations used in Mun et
al.¡¯s scheme.

\begin{table}\centering{
\caption{Notations used in the proposed scheme.}
\label{tab:1}       % Give a unique label
% For LaTeX tables use
\begin{tabular}{lllll}
\hline\noalign{\smallskip} Notation & Description\\
\hline\noalign{\smallskip}
$MU$, $FA$, $HA$& Mobile User, Foreign Agent, Home Agent\\
$PW_X$& Password of an entity $X$\\
$ID_X$& Identity of an entity $X$\\
$h(\cdot)$& A one-way hash function\\
$Cert_X$ & Certificate of an entity $X$\\
$P_X$ & Public key of $X$\\
$S_X$ & Private key of $X$\\
$E_K [\cdot]/D_K [\cdot]$ & Symmetric encryption/decryption using
key $K$\\
$E_K \{\cdot\}/D_K \{\cdot\}$ & Asymmetric encryption/decryption
using key $K$\\
$\|$& Concatenation operation\\
$\oplus$& XOR operation\\
\noalign{\smallskip}\hline
\end{tabular}}
% Or use
%\vspace*{5cm}  % with the correct table height
\end{table}

\subsection{Registration phase} \label{}
When a mobile user $MU$ wants to become a legal client to access the
services, $MU$ needs to register himself/herself to his/her home
agent $HA$.

\textbf{Step R1}: $MU$ freely chooses his/her identity $ID_{MU}$ and
password $PW_{MU}$, and generates a random number $x_{MU}$. Then
$MU$ submits $ID_{MU}$ and $h(PW_{MU}\| x_{MU})$ to $HA$ for
registration via a secure channel.

\textbf{Step R2}: When receiving the message $ID_{MU}$ and
$h(PW_{MU}\| x_{MU})$, $HA$ computes $Q=h(ID_{MU}\| y)\oplus
h(PW_{MU}\| x_{MU})$ and $H=h(ID_{MU}\| h(PW_{MU}\| x_{MU}))$. Then
$HA$ stores the message $\{Q, H, C, ID_{HA}\}$ in a smart card and
submits the smart card to $MU$ through a secure channel.

\textbf{Step R3}: After receiving the smart card, $MU$ enters
$x_{MU}$ into the smart card. Finally, $MU$'s smart card contains
parameters $\{Q, H, C, ID_{HA}, x_{MU}\}$.

The details of user registration phase are shown in Fig.4.

\begin{figure}[h]
\centering{\includegraphics[scale=0.7,trim=0 0 0 0]{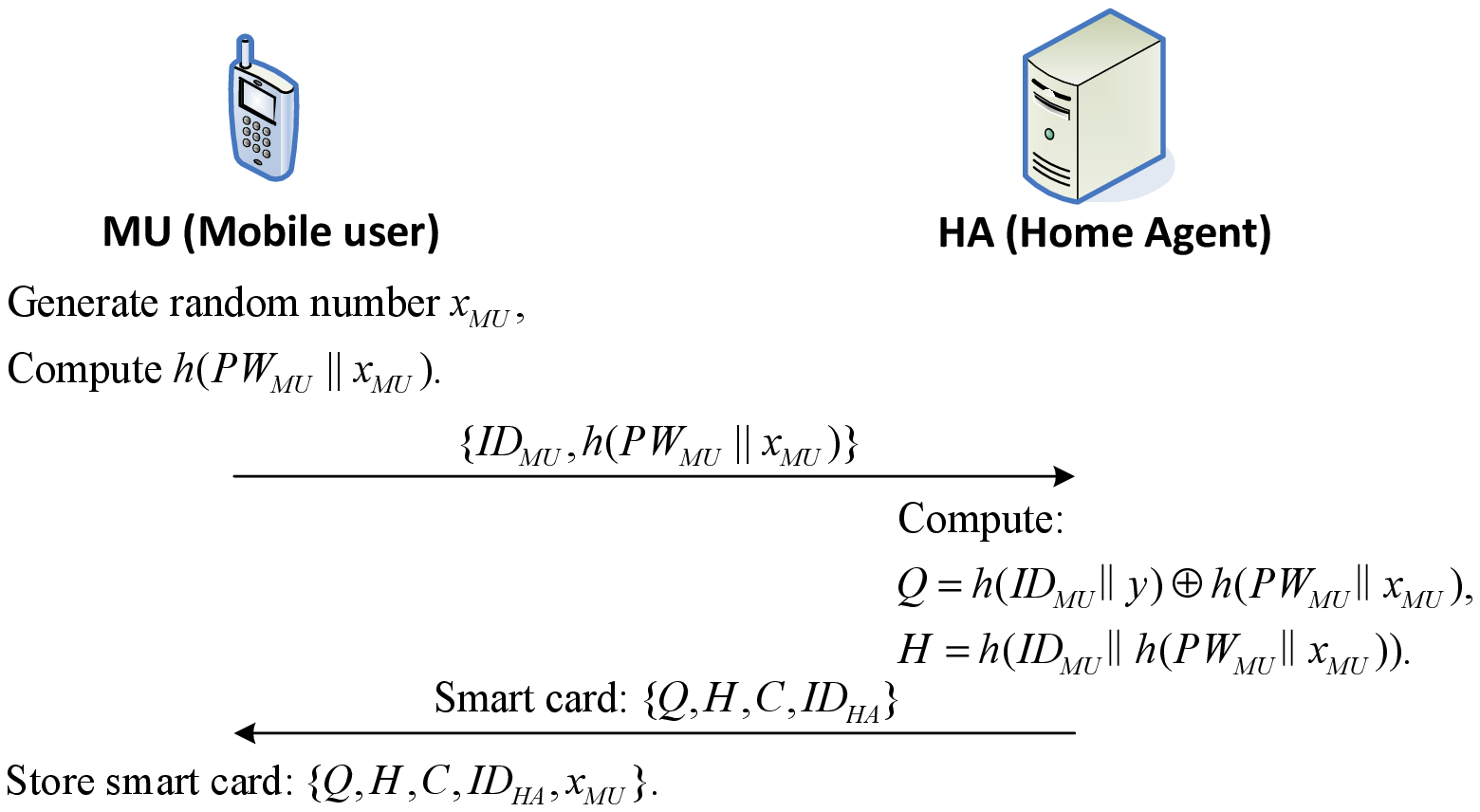}}
\caption{Registration phase of the proposed scheme.}
\end{figure}

\subsection{Authentication and establishment of session key phase} \label{}
When a mobile user $MU$ roams into a foreign network $FA$ and wants
to access services provided by $FA$. The $FA$ needs to verify the
validity of $MU$ with the assistance of $HA$, and proves to $MU$
that he is a legitimate service provider. The authentication and
establishment of session key phase of our proposed scheme is
described as follows:

\textbf{Step A1}: $MU$ inserts his/her smart card into the smart
card reader, and inputs identity $ID_{MU}$ and password $PW_{MU}$.
Then the smart card computes $H^*=h(ID_{MU}\| h(PW_{MU}\| x_{MU}))$,
and checks whether $H^*=H$. If they are equal, it means $MU$ is a
legitimate user. Otherwise the smart card aborts the session. Next,
the smart card generates a random numbers $a$, and computes $A=aP$,
$R_{AC}=aC$, $N=Q\oplus h(PW_{MU}\| x_{MU})$,
$DID_{MU}=ID_{MU}\oplus h(R_{AC})$ and $V_1=h(N\| R_{AC}\|
ID_{HA})$. Then the smart card sends the request message $\{A,
DID_{MU}, C, V_1, ID_{HA}\}$ to $FA$ over a public channel.

\textbf{Step A2}: After receiving the message $\{A, DID_{MU}, C,
V_1, ID_{HA}\}$, $FA$ generates a random numbers $b$, and computes
$B=bP$, $R_{BC}=bC$, $W_2=E_{R_{BC}}[A, Cert_{FA}, V_1, DID_{MU}]$
and $V_2=E_{S_{FA}}\{h(A, V_1,$ $DID_{MU})\}$. Here, $S_{FA}$ is the
private key of $FA$, and $Cert_{FA}$ is $FA$'s certificate. Then
$FA$ sends $\{B, W_2, V_2\}$ to $HA$.

\textbf{Step A3}: When receiving $\{B, W_2, V_2\}$, $HA$ first
computes $R_{BC}=cB$ and decrypts $D_{R_{BC}}[W_2]$ to reveal $A,
Cert_{FA}, V_1$ and $DID_{MU}$. Then, $HA$ verifies the certificate
$Cert_{FA}$ and the $FA$'s public key $P_{FA}$. If they are valid,
$HA$ verifies the $FA$'s signature $V_2$ by using the $FA$'s public
key $P_{FA}$. If they are valid, $FA$ is authenticated. After that,
$HA$ computes $R_{AC}=cA$, $ID_{MU}=DID_{MU}\oplus h(R_{AC})$ and
$V^*_1=h(h(ID_{MU}\| y)\| R_{AC}\| ID_{HA})$. Then $HA$ checks
whether $V^*_1=V_1$. If they are equal, $MU$ is authenticated by
$HA$. Next, $HA$ computes $W_1=h(h(ID_{MU}\| y)\| A\| B\| ID_{FA}\|
ID_{HA})$, $W_3=E_{R_{BC}}[ID_{FA}, Cert_{HA}, A, B, W_1]$ and
$V_3=E_{S_{HA}}\{h(Cert_{HA}, W_1)\}$. At last, $HA$ sends $\{W_3,
V_3\}$ to $FA$.

\textbf{Step A4}: $FA$ decrypts $D_{R_{BC}}[W_3]$ to reveal
$ID_{FA}, Cert_{HA}, A, B$ and $W_1$. Then, the $FA$ verifies the
$HA$'s signature $V_3$ by using the $HA$'s public key $P_{HA}$. If
it is valid, $HA$ is authenticated which also means that $HA$
claimed $MU$ is a legitimate user. After that, $FA$ computes the
common session key $SK=h(bA)$ and sends $\{B, ID_{FA}, W_1\}$ to
$MU$.

\textbf{Step A5}: After receiving the message $\{B, ID_{FA}, W_1\}$,
$MU$ computes $W^*_1=h(N\| A\| B\|$ $ID_{FA}\| ID_{HA})$ and checks
whether $W^*_1=W_1$. If they are equal, $FA$ and $HA$ are all
authenticated by $MU$. Then $MU$ establishes the common session key
$SK=h(aB)$.

The authentication and establishment of session key phase is
depicted in Fig.5.

\subsection{Update session key phase} \label{}

$MU$ and $FA$ need to renew session key for security reasons if user
is always within a same $FA$. When $MU$ visits $FA$ at the $i$th
session, the following process is conducted to authenticate $FA$:

\textbf{Step U1}: $MU$ selects a new random number $a_i$, computes
$A_i=a_iP$ $(i = 1, 2, . . . , n)$, and sends $A_i$ to FA.

\textbf{Step U2}: $FA$ selects a new random number $b_i$ and
computes $B_i=b_iP$ $(i = 1, 2, . . . , n)$. Then $FA$ generates a
new session key $SK_i= h(b_iA_i)$, and then computes $S_i =
h(b_iA_i\| SK_{i-1})$. After that, $FA$ sends $B_i$ and $S_i$ to
$MU$.

\textbf{Step U3}: $MU$ computes $S'_i= h(a_iB_i\| SK_{i-1})$ and
checks whether $S'_i=S_i$. If they are not equal, $MU$ aborts the
session. Otherwise, $MU$ computes the new session key $SK_i =
h(a_iB_i)$.

\begin{figure}[H]
\centering{\includegraphics[scale=0.7,trim=0 0 0 0]{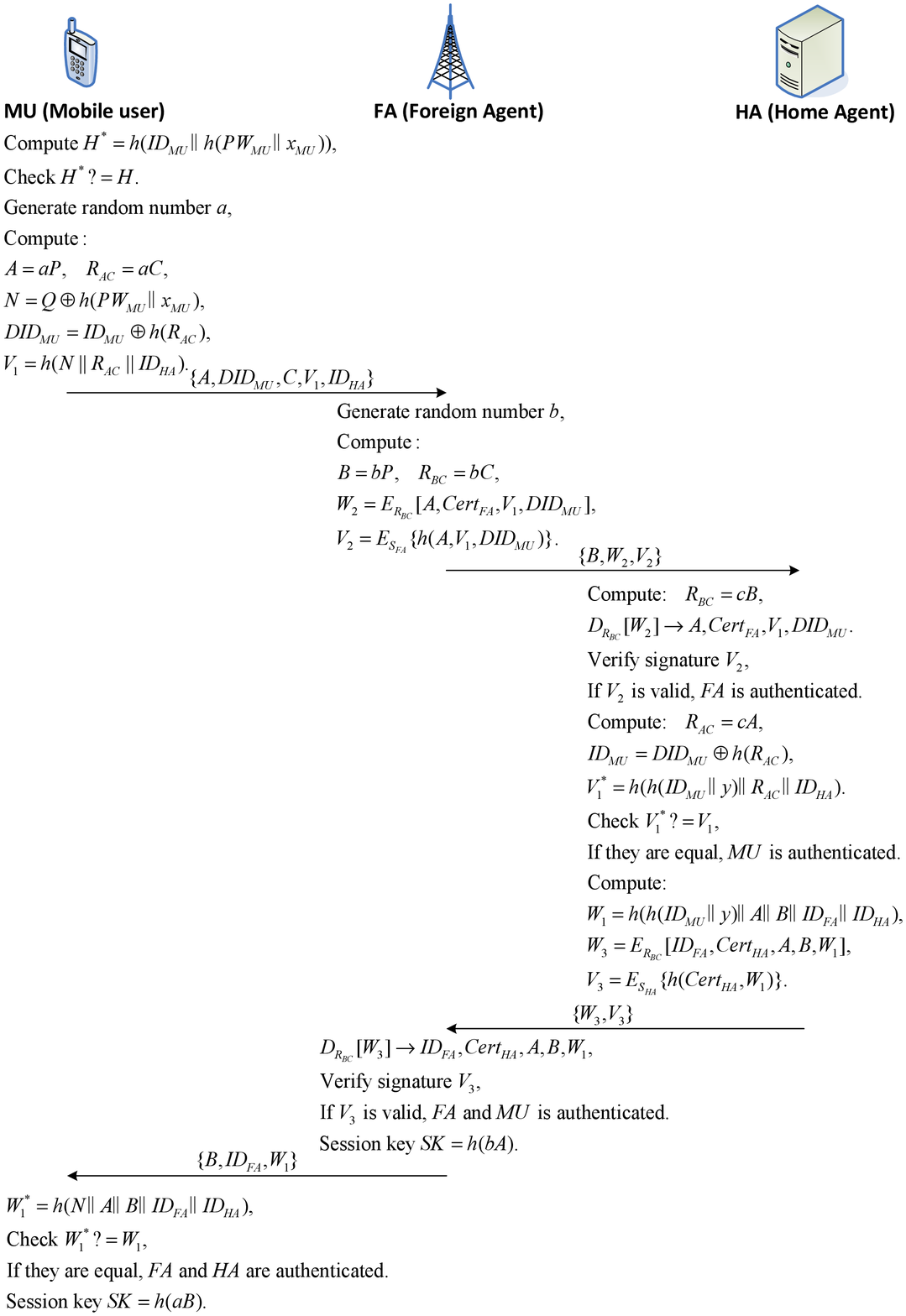}}
\caption{Authentication and establishment of session key phase of
the proposed scheme.}
\end{figure}

The details of update session key phase of the proposed scheme are
shown in Fig.6.

\begin{figure}[h]
\centering{\includegraphics[scale=0.7,trim=0 0 0 0]{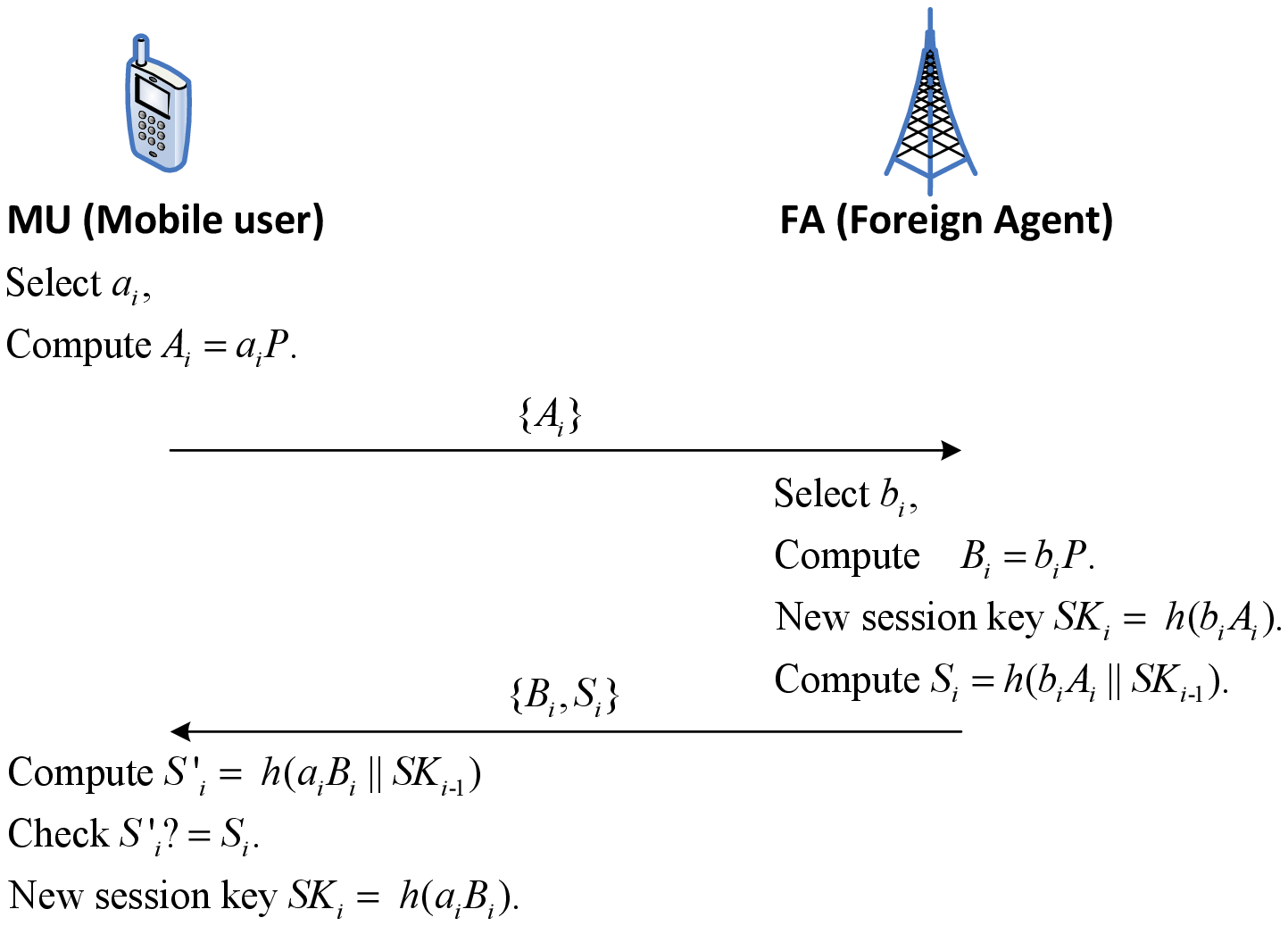}}
\caption{Update session key phase of the proposed scheme.}
\end{figure}

\subsection{Update password phase} \label{}

This phase is invoked whenever $MU$ wants to change his password
$PW_{MU}$ to a new password $PW^{new}_{MU}$. There is no need for a
secure channel for password change, and it can be finished without
communicating with his/her $HA$.

\textbf{Step U1}: $MU$ inserts his/her smart card into the smart
card reader, and inputs identity $ID_{MU}$ and password $PW_{MU}$.
Then the smart card computes $H^*=h(ID_{MU}\| h(PW_{MU}\| x_{MU}))$,
and checks whether $H^*=H$. If they are not equal, the smart card
rejects the password change request. Otherwise, $MU$ inputs a new
password $PW^{new}_{MU}$ and a new random number $x^{new}_{MU}$.

\textbf{Step U2}: The smart card computes $Q^{new}=Q \oplus
h(PW_{MU}\| x_{MU})\oplus h(PW^{new}_{MU}\| x^{new}_{MU})$ and
$H^{new}=h(ID_{MU}$ $\| h(PW^{new}_{MU}\| x^{new}_{MU}))$. Then, the
smart card replaces $Q$, $H$ and $x_{MU}$ with $Q^{new}$, $H^{new}$
and $x^{new}_{MU}$ to finish the password change phase.

\subsection{Authentication and establishment of session key scheme
when a mobile user is located in his/her home network} \label{}

Corresponding to the authentication and establishment of session key
phase when a mobile user is located in a foreign network, in this
subsection we propose an authentication and establishment of session
key scheme for that when a mobile user is located in his/her home
network. The detail processes are described as follows and depicted
in Fig.7.

\begin{figure}[H]
\centering{\includegraphics[scale=0.7,trim=0 0 0 0]{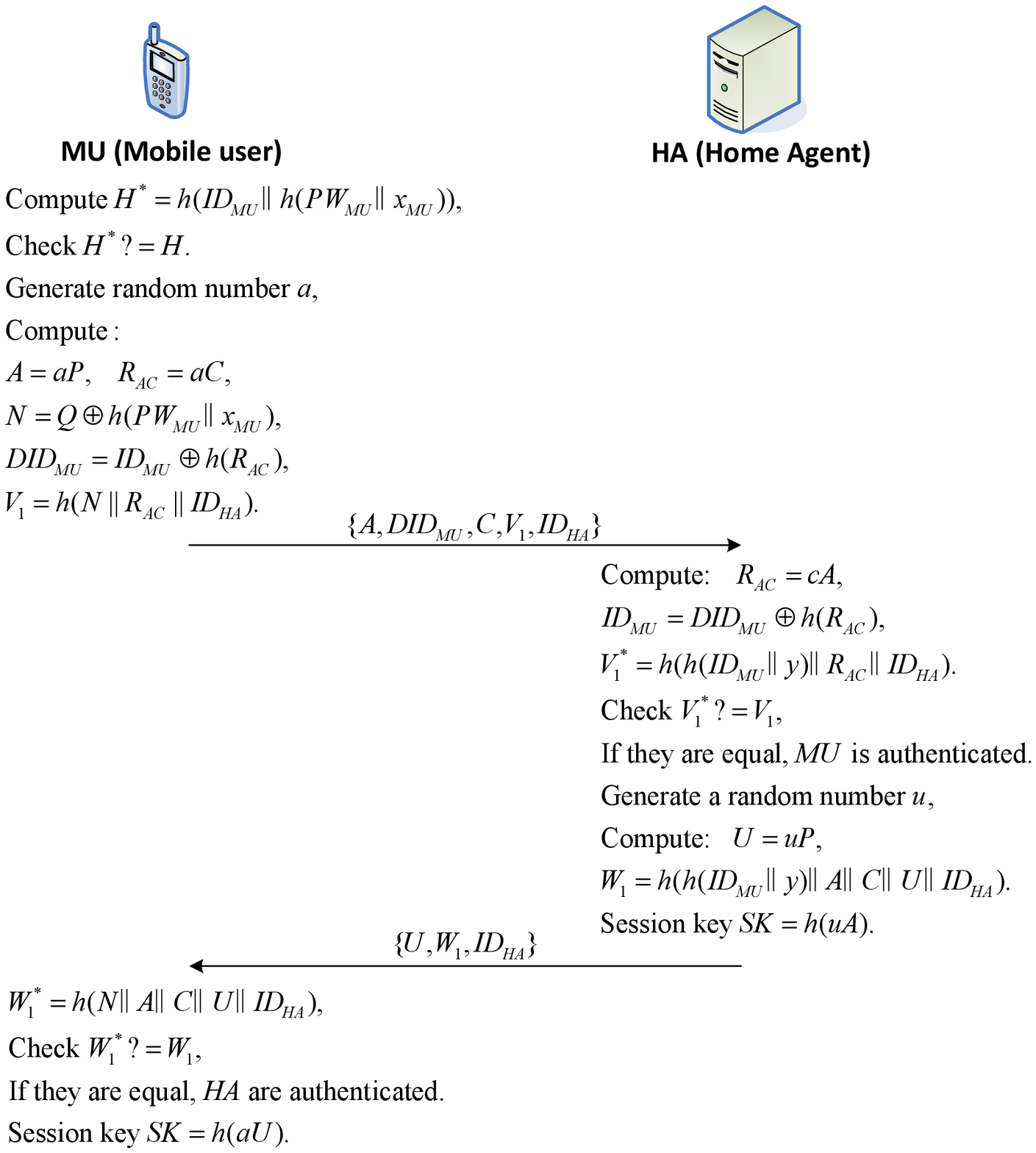}}
\caption{Authentication and establishment of session key scheme when
a mobile user is located in his/her home network.}
\end{figure}

\textbf{Step A1}: $MU$ inserts his/her smart card into the smart
card reader, and inputs identity $ID_{MU}$ and password $PW_{MU}$.
Then the smart card computes $H^*=h(ID_{MU}\| h(PW_{MU}\| x_{MU}))$,
and checks whether $H^*=H$. If they are equal, it means $MU$ is a
legitimate user. Otherwise the smart card aborts the session. Next,
the smart card generates a random numbers $a$, and computes $A=aP$,
$R_{AC}=aC$, $N=Q\oplus h(PW_{MU}\| x_{MU})$,
$DID_{MU}=ID_{MU}\oplus h(R_{AC})$ and $V_1=h(N\| R_{AC}\|
ID_{HA})$. Then the smart card sends the request message $\{A,
DID_{MU}, C, V_1, ID_{HA}\}$ to $HA$ over a public channel.

\textbf{Step A2}: After receiving the message $\{A, DID_{MU}, C,
V_1, ID_{HA}\}$, $HA$ first computes $R_{AC}=cA$ and
$ID_{MU}=DID_{MU}\oplus h(R_{AC})$ and $V^*_1=h(h(ID_{MU}\| y)\|
R_{AC}\| ID_{HA})$. Then $HA$ checks whether $V^*_1=V_1$. If they
are equal, $MU$ is authenticated by $HA$. Next, $HA$ generates a
random number $u$, and computes $U=uP$ and $W_1=h(h(ID_{MU}\| y)\|
A\| C\| U\| ID_{HA})$. At last, $HA$ computes the session key
$SK=h(uA)$ and sends $\{U, W_1, ID_{HA}\}$ to $MU$.

\textbf{Step A3}: When receiving the message $\{U, W_1, ID_{HA}\}$,
$MU$ computes $W^*_1=h(N\| A\| C\| U\|$ $ID_{HA})$ and checks
whether $W^*_1=W_1$. If they are equal, $HA$ is authenticated by
$MU$. Then $MU$ establishes the common session key $SK=h(aU)$.

\section{Security analysis of the proposed scheme} \label{}

In this section, we show that the proposed scheme can withstand all
possible security attacks and can work correctly.

\textbf{Proposition 1.} The proposed scheme can provide user's
anonymity.

\textbf{Proof.} In our proposed scheme, the mobile user $MU$ sends
the login request message $\{A,$ $DID_{MU}, C,$ $V_1, ID_{HA}\}$ to
$FA$, where $DID_{MU}=ID_{MU}\oplus h(aC)$ is used to protect the
real identity $ID_{MU}$ of $MU$. Based on the CDL problem, any
attacker cannot obtain the random number $a$ form $A$ and thus
cannot retrieve $ID_{MU}$ from $DID_{MU}$. At the same time, the
attacker cannot trace the moving history and current location of
$MU$ according to the login request message since $A$, $DID_{MU}$
and $V_1$ are dynamically changed in different login request
messages of $MU$. Therefore, the proposed scheme can provide user's
anonymity.

\textbf{Proposition 2.} The proposed scheme can provide proper
mutual authentication and thus prevent impersonation attack.

\textbf{Proof.} In order to impersonation attack, the mobile user
$MU$, the foreign agent $FA$, and the home agent $HA$ should
authenticate each other, which requires that our protocol provides
mutual authentication mechanism between any two of them. The
proposed scheme can efficiently prevent impersonation attacks by
considering the following scenarios:

(1) The proposed scheme provides authentication of $FA$ and $HA$ to
$MU$, and thus attacker cannot impersonate $MU$ to cheat $FA$ and
$HA$. In the proposed scheme, whether MU is located in a foreign
network or in his/her home network, the $HA$ authenticates $MU$ by
verifying the computed $V^*_1=h(h(ID_{MU}\| y)\| R_{AC}\| ID_{HA})$
with the received $V_1=h(N\| R_{AC}\| ID_{HA})$. Since the attacker
does not possess $MU$'s password $PW_{MU}$, he/she cannot compute
the correct $N=Q\oplus h(PW_{MU}\| x_{MU})$ and thus cannot cheat
$HA$ by forging a login request message. At the same time, since $a$
is a one-time random number and only possessed by $MU$, $V_1$ is
dynamically changed in each login request message. Therefore, the
attacker cannot cheat the $HA$ by replaying a previous login request
message. Beside, when MU is located in a foreign network, the
authentication of $FA$ to $MU$ is completely dependent on the
authentication of $HA$ to $MU$. If an attacker cannot successfully
cheat $HA$ by masquerading as $MU$, he/she cannot cheat $FA$
successfully.

(2) The proposed scheme provides authentication of $HA$ and $MU$ to
$FA$, and thus attacker cannot impersonate $FA$ to cheat $HA$ and
$MU$. In the proposed scheme, the $HA$ authenticates $FA$ by
checking whether $D_{P_{FA}}\{V_2\}$ equals $h(A, V_1,$ $DID_{MU})$,
where $V_2$ is $FA$'s digital signature. Obviously, the attacker
cannot compute the correct $FA$'s digital signature without knowing
$FA$'s private key $S_{FA}$. Therefore, the attacker cannot cheat
$HA$ successfully by masquerading as $FA$. At the same time, the
authentication of $MU$ to $FA$ is completely dependent on the
authentication of $HA$ to $FA$. If an attacker cannot successfully
cheat $HA$ by masquerading as $FA$, he/she cannot cheat $MU$
successfully.

(3) The proposed scheme provides authentication of $FA$ and $MU$ to
$HA$, and thus attacker cannot impersonate $HA$ to cheat $FA$ and
$MU$. In the proposed scheme, the $FA$ authenticates $HA$ by
checking whether $D_{P_{HA}}\{V_3\}$ equals $h(Cert_{HA}, W_1)$,
where $V_3$ is $HA$'s digital signature. Obviously, the attacker
cannot compute the correct $HA$'s digital signature without knowing
$HA$'s private key $S_{HA}$. Therefore, the attacker cannot cheat
$FA$ successfully by masquerading as $HA$. Besides, the $MU$
authenticates $HA$ by verifying the computed $W^*_1=h(N\| A\| B\|
ID_{FA}\| ID_{HA})$ with the received $W_1=h(h(ID_{MU}\| y)\| A\|
B\| ID_{FA}\| ID_{HA})$. Since any attacker cannot compute the
correct $W_1$ without knowing $ID_{MU}$ and $y$, the attacker cannot
cheat $MU$ successfully.

\textbf{Proposition 3.} The proposed scheme can withstand the replay
attack.

\textbf{Proof.} An attacker might replay an old login request
message $\{A, DID_{MU}, C, V_1, ID_{HA}\}$ to $FA$ and receive the
message $\{B, ID_{FA}, W_1\}$ from $FA$. However, the attacker still
cannot compute the correct session key $SK=h(aB)$ since he/she
cannot derive the secret information $a$ form $A=aP$ based on the
security of CDL problem. Thus, the proposed scheme can prevent the
replay attack.

\textbf{Proposition 4.} The proposed scheme meets the security
requirement for perfect forward secrecy.

\textbf{Proof.} Perfect forward secrecy means that even if an
attacker compromises all the passwords of the entities of the
system, he/her still cannot compromise the session key. In the
proposed scheme, the session key $SK=h(abP)$ is generated by two
one-time random numbers $a$ and $b$ in each session. These two
one-time random numbers are only held by the $MU$ and $FA$
respectively, and cannot be retrieved from $A=aP$, $B=bP$,
$R_{AC}=aC=cA$ and $R_{BC}=bC=cB$ based on the security of CDL and
CDH problem. Thus, even if an adversary obtains all the passwords of
the entities, previous session keys and all the transmitted
messages, he/her still cannot compromise other session key. Hence,
the proposed scheme achieves perfect forward secrecy.

\textbf{Proposition 5.} Our scheme can resist off-line password
guessing attack with smart card security breach.

\textbf{Proof.} In the proposed scheme, it is assume that if a smart
card is stolen, physical protection methods cannot prevent malicious
attackers to get the stored secure elements. At the same time,
attacker can access to a big dictionary of words that likely
includes user's password and intercept the communications between
the user and server.

It is assumed that an attacker has obtained the information $\{Q, H,
C, ID_{HA}, x_{MU}\}$ from the stolen $MU$'s smart card and has
intercepted a previous full transmitted messages $\{A, DID_{MU}, C,$
$V_1, ID_{HA},$ $B, W_2, V_2, W_3, V_3, B, ID_{FA}, W_1\}$. In the
proposed scheme, $MU$'s password only makes two appearances as
$H=h(ID_{MU}\| h(PW_{MU}\| x_{MU}))$ and $V_1=h((Q\oplus h(PW_{MU}\|
x_{MU}))\|$ $aC \| ID_{HA})$. Obviously, the attacker cannot launch
an off-line password guessing attack without knowing the $ID_{MU}$
and $a$. Since it has been demonstrated that our scheme can provide
user anonymity and $a$ is $MU$'s secret random number, the proposed
scheme can resist off-line password guessing attack with smart card
security breach.

\textbf{Proposition 6.} The proposed scheme can withstand insider
attack.

\textbf{Proof.} If an insider of the home agent $HA$ has obtained a
user $MU$'s password $PW_{MU}$, he/she can impersonate as $MU$ to
access any foreign agent. In the registration phase of the proposed
scheme, $MU$ sends identity $ID_{MU}$ and $h(PW_{MU}\| x_{MU})$ to
$HA$. Thus, the insider cannot derive $PW_{MU}$ without $x_{MU}$.
Besides, in the password change phase, $MU$ can change his/her
default password $PW_{MU}$ without the assistance of his/her $HA$.
Therefore the insider has no chance to obtain $MU$'s password, our
scheme can withstand the insider attack.

\textbf{Proposition 7.} There is no verification table in the
proposed scheme.

\textbf{Proof.} In the proposed scheme, it is obvious that the user,
the foreign agent and the home agent do not maintain any
verification table.

\textbf{Proposition 8.} The proposed scheme can provide local
password verification.

\textbf{Proof.} In the proposed scheme, smart card checks the
validity of $MU$'s identity $ID_{MU}$ and password $PW_{MU}$ before
logging into $FA$. Since the attacker cannot compute the correct $H$
without the knowledge of $ID_{MU}$ and $PW_{MU}$ to pass the
verification equation $H^*=H$, thus our scheme can avoid the
unauthorized accessing by the local password verification.

\section{Performance comparison and functionality analysis}
\label{}

In this section, we compares the performance and functionality of
our proposed scheme with some previously schemes. It is well-known
that most of the mobile devices have limited energy resources and
computing capability. Hence, one of the most important issues in
wireless networks is power consumption caused by communication and
computation. In fact, the communication cost in the GLOMONET is
higher than computation cost in terms of power consumption. In table
3, we list the numbers of the message exchanges in the login,
authentication and session key establish phases of our scheme and
some related previous schemes. And the bit-length of communication
of the mobile client in these phases is also shown since the foreign
agent and home agent are regarded as powerful devices. Table 4 shows
the computational cost of our proposed scheme and some other related
protocols. Here we mainly focus on the computational cost of the
login, authentication and session key establish phases because these
phases are the principal part of an authentication scheme. In
general, our proposed scheme spends relatively few communication and
computational cost. It is suitable for the low-power and
resource-limited mobile devices.

\begin{table}
\caption{Communication  cost comparison of our scheme and other
schemes.}
\label{tab:1}       % Give a unique label
% For LaTeX tables use
\begin{tabular}{lllllllllllll}
\hline\noalign{\smallskip}
& Our scheme & He et al. [2] & Li et al. [7] & Mun et al. [8]\\
\hline\noalign{\smallskip}
Communication (bits) & 3808 & 2240 & 8224 & 4192\\
Communication (rounds) & 4 & 4 & 4 & 5\\
\noalign{\smallskip}\hline
\end{tabular}\\
The bit-length of different parameter: $xP$: 1024, $g^x$ mod$p$:
1024, identity $ID_{x}$: 160, time: 128, random number: 128, hash
function $h(x)$: 160, encryption/decryption: 1024.
\end{table}

\begin{table}\footnotesize
\caption{Computational cost comparison of our scheme and other
schemes.}
\label{tab:1}       % Give a unique label
% For LaTeX tables use
\begin{tabular}{lllllllllllll}
\hline\noalign{\smallskip}
& & $Add$ & $Hash$ & $Mod$ & $Mul$ & $Esym$ & $Dsym$ & $Easym$ & $Dasym$ & $Gsign$ & $Vsign$\\
\hline\noalign{\smallskip}
& MU &2&6&N/A&1+2Pre&N/A&N/A&N/A&N/A&N/A&N/A\\
Our scheme & FA &N/A&1&N/A&2+Pre&1&1&N/A&N/A&1&1\\
& HA &1&4&N/A&2&1&1&N/A&N/A&1&1\\
\hline\noalign{\smallskip}
& MU &5&10&N/A&N/A&1&1&N/A&N/A&N/A&N/A\\
He et al. [2] & FA &N/A&2&N/A&N/A&1&N/A&N/A&1&1&1\\
& HA &2&3&N/A&N/A&N/A&2&1&N/A&1&1\\
\hline\noalign{\smallskip}
& MU &4&2&1+3Pre&N/A&3&1&N/A&N/A&N/A&N/A\\
Li-Lee [7] & FA &N/A&1&3+2Pre&N/A&2&2&N/A&N/A&1&1\\
& HA &2&3&3+Pre&N/A&1&3&N/A&N/A&1&1\\
\hline\noalign{\smallskip}
& MU &2&4&N/A&1+Pre&1&N/A&N/A&N/A&N/A&N/A\\
Mun et al. [8] & FA &2&3&N/A&1+Pre&1&N/A&N/A&N/A&N/A&N/A\\
& HA &3&3&N/A&N/A&N/A&N/A&N/A&N/A&N/A&N/A\\
\noalign{\smallskip}\hline
\end{tabular}
Note: Pre: pre-computed operation, $Add$: XOR operation, $Hash$:
hash operation, $Mod$: modular exponentiation, $Mul$: point scalar
multiplication, $Esym$: Symmetric encryption $E_K[\cdot]$, $Dsym$:
Symmetric decryption $D_K[\cdot]$, $Easym$: Asymmetric encryption
$E_K\{\cdot\}$, $Dasym$: Asymmetric decryption $D_K\{\cdot\}$,
$Gsign$: Signature generation $E_K\{h(\cdot)\}$, $Vsign$: Signature
verification $D_K\{h(\cdot)\}$.
\end{table}

Table 5 lists the functionality comparisons among our proposed
scheme and other related schemes. It is obviously that our scheme
has many excellent features and is more secure than other related
schemes.

\begin{table}\centering{
\caption{Functionality comparison between the related schemes and
our scheme.}
\label{tab:1}       % Give a unique label
% For LaTeX tables use
\begin{tabular}{llllllllll}
\hline\noalign{\smallskip}
  & Our & Wu & Chang & He & He & Mun & Li\\
  & scheme & et al. & et al. & et al. & et al. & et al. & et al.\\
%  &  & (2008) & (2009) & (2011) & (2011) & (2012) & (2012)\\
  &  & [6] & [5] & [2] & [9] & [8] & [7]\\
\hline\noalign{\smallskip}
User's anonymity &Yes& No & No & No & No & No & Yes\\
Proper mutual authentication &Yes& No & Yes & Yes & No & No & Yes\\
Resist $MU$ impersonation attack &Yes& No &Yes & Yes & No & No & Yes\\
Resist $FA$ impersonation attack &Yes& Yes &Yes & Yes & Yes & No & Yes\\
Resist $HA$ impersonation attack &Yes& Yes &Yes & Yes & Yes & No & Yes\\
Resist replay attack &Yes& No & Yes & Yes & No & No & No\\
Perfect forward secrecy &Yes& No & No & No & No & Yes & Yes\\
Resist off-line password guessing attack &Yes& No & No & Yes & No & No  & Yes\\
Resist insider attack &Yes& No & No & Yes & No & No & Yes\\
No verification table &Yes&  Yes  & No & Yes & No & Yes & Yes\\
Local password verification &Yes& No & No & Yes & Yes & No & No\\
Correct password change &Yes & No & No & Yes & No & No & Yes\\
Provid the authentication scheme when\\
user is located in his/her home network &Yes & No & No & Yes & No & No & No\\
\noalign{\smallskip}\hline
\end{tabular}}
% Or use
%\vspace*{5cm}  % with the correct table height
\end{table}

\section{Conclusion}

In this paper, we show that the recently proposed Mun et al.'s
authentication scheme for roaming service cannot provide user
friendliness, user's anonymity, proper mutual authentication and
local verification and also vulnerable to several attacks. In order
to withstand security flaws in Mun et al.'s scheme, we propose a
novel anonymous authentication scheme for roaming service in global
mobility networks. Security and performance analyses show the
proposed scheme is more suitable for the low-power and
resource-limited mobile devices, and is secure against various
attacks and has many excellent features.

\section{Acknowledgment}

This paper was supported by the National Natural Science Foundation
of China (Grant Nos. 61070209, 61202362, 61121061), and the Asia
Foresight Program under NSFC Grant  (Grant No. 61161140320).

\end{document}